%% file: El_embrollo_de_Bernoulli.tex
\documentclass[10pt]{book}

\input{msel}

\usepackage[spanish]{babel}
\usepackage{wrapfig}
\usepackage{graphicx}
\usepackage{caption}
\usepackage{subcaption}
\usepackage{color}

\begin{document}

\renewcommand{\bibname}{R\MakeLowercase{eferencias}}
\renewcommand{\tablename}{Tabla}
\renewcommand{\figurename}{Figura}

\renewcommand{\thetable}{\arabic{table}}
\renewcommand{\theequation}{\arabic{equation}}
\renewcommand{\thefigure}{\arabic{figure}}

\def\citeapos#1{\citeauthor{#1} \citeyear{#1}}


\normalsize
\chapter[El Embrollo de Bernoulli \\
\color{black} {\sc Suárez Á., Dutra M., Monteiro M., Martí
  A.}]{Bernoulli’s muddle: a research on students’ misconceptions in
  fluid dynamics\\ El embrollo de Bernoulli: una investigación sobre
  las concepciones alternativas de los estudiantes en dinámica de
  fluidos } \minitoc


\renewcommand\author1{
\begin{minipage}{.4\textwidth}
{\bf  Álvaro Suárez} \\ 
{\sc Consejo de Formación en Educación, Montevideo, Uruguay.} \\ 
{\href{mailto:alsua@outlook.com}{alsua2000@gmail.com}}
 \end{minipage}}
 \hfill \renewcommand\author2{\begin{minipage}{.4\textwidth} {\bf
       Mateo Dutra} \\ {\sc Facultad de Ciencias, Universidad de la
       República, Montevideo, Uruguay.}
     \\ {\href{mailto:mateodutrafisica@gmail.com}{mateodutrafisica@gmail.com}}
\end{minipage}}
 \hfill
\renewcommand\author3{
\begin{minipage}{.4\textwidth}
{\bf  Martín Monteiro} \\ 
{\sc Universidad ORT Uruguay, Uruguay.} \\ 
{\href{mailto:fisica.martin@gmail.com}{fisica.martin@gmail.com}}
 \end{minipage}}
 \hfill \renewcommand\author4{\begin{minipage}{.4\textwidth} {\bf
       Arturo C. Marti} \\ {\sc Facultad de Ciencias, Universidad de
       la República, Montevideo, Uruguay.}
     \\ {\href{mailto:marti@fisica.edu.uy}{marti@fisica.edu.uy}}
\end{minipage}}

\abst{\large Bernoulli's equation, which relates the pressure of an
  ideal fluid in motion with its velocity and height under certain
  conditions, is a central topic in General Physics courses for
  Science and Engineering students. This equation, frequently used
  both textbooks as in science outreach activities or museums, is
  often extrapolated to explain situations in which it is no longer
  valid. A common example is to assume that, in any situation, higher
  speed means lower pressure, a conclusion that is only acceptable
  under certain conditions. In this paper we report the results of an
  investigation with university students on some misconceptions
  present in fluid dynamics. We found that after completing the
  General Physics courses, many students have not developed a correct
  model about the interaction of a fluid element with its environment
  and extrapolate the idea that higher speed implies lower pressure in
  situations where it is no longer valid. We also show that an
  approach to fluid dynamics based on Newton's laws is more natural to
  address these misconceptions.  \\[2mm] La ecuación de Bernoulli, que
  bajo ciertas condiciones relaciona la presión de un fluido ideal en
  movimiento con su velocidad y su altura, es un tema central en los
  cursos de Física General para estudiantes de Ciencias e
  Ingeniería. Frecuentemente, en los libros de texto utilizados en
  cursos universitarios, al igual que en diversos medios de
  divulgación, se suele extrapolar este principio para explicar
  situaciones en las que no es válido. Un ejemplo habitual es suponer
  que, en cualquier situación, mayor velocidad implica menor presión,
  conclusión correcta solo en algunas circunstancias. En este trabajo
  presentamos los resultados de una investigación con
  estudiantes universitarios, sobre las concepciones alternativas
  presentes en dinámica de fluidos. Encontramos que muchos
  estudiantes, incluso después de haber transitado por los cursos de
  Física General, no han elaborado un modelo adecuado acerca de la
  interacción de un elemento de un fluido con su entorno y extrapolan
  la idea que mayor velocidad implica una menor presión en contextos
  donde no es válida. Mostramos también que un enfoque de la dinámica
  de fluidos basado en las leyes de Newton resulta más natural para
  confrontar estas concepciones alternativas.}

\noindent \keyword{Bernoulli;fluid dynamics, misconceptions}

\noindent \clave{Bernoulli; dinámica de fluidos, concepciones alternativas }
\newpage
\large


\section{Introducción}

Todas las personas curiosas se han preguntado alguna vez por qué
vuelan los aviones o por qué un balón de fútbol se desvía lateralmente
cuando se lo hace rotar al ser pateado (fenómeno conocido como
\textit{patear con efecto, chanfle, o comba} en los países
iberoamericanos). Las explicaciones a estas interrogantes encontradas
en muchos libros, artículos, conferencias, museos y sitios web además
de ser simples y elegantes, suelen tener en común la idea que, en un
fluido, cuanto mayor es la velocidad (en módulo o valor absoluto)
menor es la presión. Asimismo, frecuentemente se proponen experimentos
caseros, o se muestran vídeos, tales como soplar por encima de un
papel o colocar una pelota de \textit{ping pong} (tenis de mesa) por
encima de un secador de pelo, que confirmarían esta relación
\cite{barbosa2013montajes,ehrlich1990turning,pedros2013demo}. Si bien
en muchos casos la explicación basada en la idea menor presión-mayor
velocidad es correcta en muchos otros conduce a contradicciones con
los experimentos. Esto no debería sorprendernos dado
  que la ecuación de Bernoulli es válida bajo ciertas hipótesis, en
  particular que los puntos donde se aplica deben estar en la misma
  línea de corriente, condición que en varios de los ejemplos
  propuestos no se cumple.

La idea que cuanto mayor es la velocidad de un fluido su presión
siempre disminuye está tan extendida, que aparece como explicación de
diversos fenómenos señalados en museos de ciencias y material de
divulgación de agencias mundialmente conocidas. Por ejemplo, en
material de la NASA se proponen experimentos que involucran este tipo
de fenómenos para comprender la ecuación de Bernoulli
\cite{gipson2017principles}, o en el museo ``LIGO's Science Education
Center'' del Caltech, en Livingston, Louisiana, EEUU, se la utiliza
para explicar el vuelo de los aviones con frases como ``Esto es debido
al principio de Bernoulli que dice que a mayor velocidad de un gas o
un líquido, menor es la presión resultante''.  Esta idea ha permeado la cultura popular convirtiéndose en  obstáculo para el aprendizaje y fuente de errores conceptuales en los estudiantes.

En enseñanza de la ciencia, los errores conceptuales
se caracterizan por tener un origen común y estar firmemente arraigados en  estudiantes que los cometen convencidos de su veracidad, sin manifestar dudas y a lo largo de diferentes niveles educativos y ubicaciones geográficas \cite{carrascosa2005problema}. Frecuentemente al comenzar las clases de ciencias, los alumnos llevan años interactuando e interpretando el mundo que los rodea a partir de la información que proviene de sus sentidos y desarrollando sus propias ideas para explicar los diferentes fenómenos observados. Estos modelos, aunque no conformen una teoría e incluso en     muchas circunstancias resulten incoherentes, logran predecir exitosámente algunas situaciones. De esta forma los estudiantes frecuentemente elaboran sus propias  explicaciones \textit{ad hoc} para distintos fenómenos observados. El tiempo transcurrido y su capacidad predictiva, aún bajo un marco limitado y contradictorio, conduce a  que estas ideas resulten muy persistentes y resistentes al cambio incluso al transistar por  el sistema educativo, donde interfieren  con las concepciones científicas que se busca enseñar \cite{pozo1991procesos}. 
Estas ideas  se denominan en la literatura \textit{concepciones alternativas}\footnote{También se las denomina en la literatura como preconceptos, ideas previas, concepciones erróneas o concepciones espontáneas. Aunque estos términos no son estrictamente sinónimos, ya que cada uno descansa en una posición epistemológica diferente, en todos los casos refieren a  modelos de pensamiento de los estudiantes surgidos de su interacción con el mundo que los rodea \cite{pozo1991procesos}.}
y son el origen de los errores conceptuales. Su conocimiento en las diferentes áreas de la Física, es clave para el desarrollo de actividades tendientes a modificarlas, ya que para que un estudiante remplace una concepción alternativa por la idea científicamente correcta necesita presentar cierto grado de insatisfacción con sus propias ideas 
\cite{carey1999conceptual}.

Aunque en mucho menor extensión que en otras áreas de la Física,
en cinemática y dinámica de los fluidos se han desarrollado investigaciones sobre errores conceptuales y concepciones alternativas. En esa línea,  \citeapos{suarez2017students} basados en el análisis de entrevistas transmiten que muchos estudiantes suponen que siempre que la velocidad de un fluido aumenta, su presión disminuye independientemente de si el fluido está o no confinado. Comunican también que muchos estudiantes
tienen grandes dificultades para reconocer la manera en que interactúa
un elemento de volumen de un fluido en movimiento con su entorno.  En
algunos casos consideran que un fluido confinado tiene un
comportamiento similar al de un conjunto de partículas que no
interaccionan entre sí; hallazgo confirmado recientemente por
\citeapos{schafle2019students}; mientras que en otros, suponen que un
elemento de volumen en movimiento es afectado solamente por el fluido
que lo precede. Estas investigaciones sugieren que la idea que mayor
velocidad siempre implica menor presión puede tener también una raíz
en un pobre entendimiento de la dinámica de los fluidos.
Denominaremos en este trabajo
``el embrollo de Bernoulli'' a la  generalización excesiva de  la relación entre velocidad y presión en un fluido incompresible.

Es interesante notar que la concepción alternativa mayor velocidad-
menor presión, producto de la educación formal y no formal de nuestros
estudiantes, compite con la idea previa originada de la experiencia
cotidiana que cuanto mayor es la presión de un fluido, mayor es su
velocidad. Esta idea previa ha sido ampliamente discutida en la
literatura
\cite{barbosa2013construccion,martin1983misunderstanding,vega2017dificultades}
y podría tener diversos
orígenes. \citeapos{martin1983misunderstanding}, argumenta que las
dificultades de los estudiantes están asociadas a cómo interpretan el
experimento cotidiano de apretar con el dedo la boca de salida de una
manguera. Cualquiera que haya utilizado dicho artilugio para aumentar
la velocidad de salida de un fluido de una manguera o una canilla,
reconoce que la fuerza ejercida por el agua sobre el dedo es mayor,
cuanto más pequeña sea la abertura que se deje. Este hecho
experimental, junto al asociar la fuerza sobre el dedo con la presión,
conlleva a los estudiantes a concebir que cuanto mayor sea la
velocidad, mayor debe ser la presión. Según
\citeapos{barbosa2013construccion}, la concepción de los estudiantes
que los lleva a concluir que mayor presión implica mayor velocidad,
está asociado con la creencia que la presión equivale a la fuerza y a
su vez vincular la fuerza con la velocidad. Para
\citeapos{vega2017dificultades}, tiene su origen en una idea previa de
relacionar la presión con el espacio ocupado por el fluido
\cite{besson2004students,goszewski2013exploring}, lo que lleva a
suponer que la presión aumenta en los lugares estrechos y como
consecuencia de ello, la velocidad es mayor.

Del análisis de la literatura se desprende que, mientras existe vasta
evidencia del impacto de las ideas previas derivadas de la experiencia
cotidiana al relacionar la velocidad de un fluido con su presión, es
necesario continuar indagando sobre la manera en que los estudiantes
transfieren las conclusiones derivadas de la ecuación de Bernoulli a
escenarios donde no son válidas. Las investigaciones realizadas sobre
este punto han sido de corte cualitativo con muestras pequeñas de
estudiantes. En este sentido para cuantificar el impacto de estas
concepciones alternativas resulta necesario realizar investigaciones
más amplias. Con este objetivo, en este trabajo describimos una
investigación con estudiantes universitarios sobre el impacto que
tienen las aplicaciones erróneas de la ecuación de
  Bernoulli y en particular la extrapolación de la idea mayor
velocidad-menor presión en contextos donde no es válida. En lo que
sigue, el trabajo está estructurado de la siguiente manera: en la
siguiente sección describimos cómo se llega a la idea de mayor
velocidad-menor presión a partir de la ecuación de Bernoulli y hacemos
una reseña de varios experimentos que aparecen en la literatura que no
pueden ser explicados a partir de dicha premisa. A continuación, en la
sección 3 describimos la metodología de la investigación y en la
sección 4 los resultados. Finalmente, en la sección 5 presentamos las
consideraciones finales.

\section{El embrollo de Bernoulli}
La ecuación de Bernoulli, en su versión más simple, surge de aplicar
el teorema del trabajo y la energía cinética a un fluido
incompresible, no viscoso, irrotacional, estacionario y confinado (ver
por ejemplo \citeapos{resnick2002physics}). Esta ecuación que relaciona las energías cinética y potencial  gravitatoria con los trabajos realizados por las presiones en las  fronteras de un volumen de control se expresa habitualmente como
\begin{equation}
    P + \rho gh + \frac{\rho v^2}{2} = \mathrm{constante}
    \label{eq1}
\end{equation}
en puntos sobre una misma línea de corriente donde $P$ es la presión,
$\rho$ la densidad, $h$ la altura, $v$ la velocidad y $g$ la
aceleración gravitatoria.

\begin{wrapfigure}{r}{0.35\textwidth}
    \includegraphics[width=0.25\textwidth]{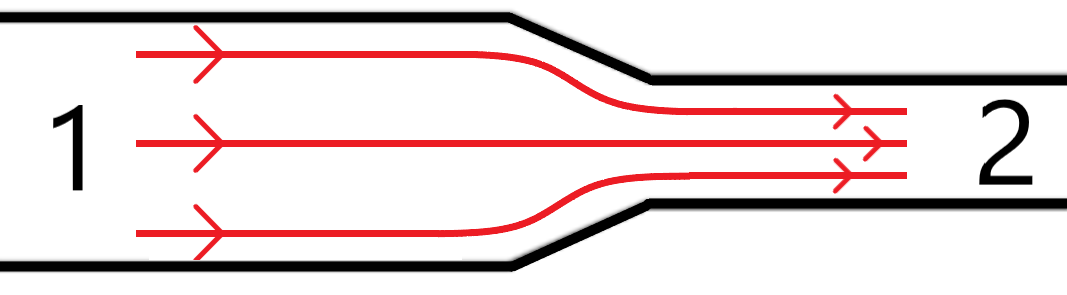}
    \caption{Fluido en una tubería horizontal. Dado que se verifican
      las hipótesis necesarias al aplicar la ecuación de Bernoulli
      deducimos que la presión del fluido  en el punto~1 es mayor que
      en el 2.}
    \label{fig1}
\end{wrapfigure}

Resulta llamativo que la ecuación de Bernoulli
deducida bajo hipótesis restrictivas se presenta en los textos de Física General   \cite{resnick2002physics,sears2013fisica,jewett2008fisica,tipler2004fisica} como un principio fundamental de la dinámica de fluidos. En efecto,
el vasto conjunto de fenómenos donde se muestra su aplicación  sugiere un rol mucho más importante del real en el marco general de
la hidrodinámica. En este trabajo discutimos si este exceso de
jerarquía en la presentación contribuye a potenciar las posibles
dificultades conceptuales que aparecen en el entendimiento de la
temática.

Analizar detalladamente la literatura de referencia y los libros de
textos de Física General donde se presenta la ecuación de Bernoulli
nos revela algunas pistas sobre el embrollo de Bernoulli. Cuando
aplicamos esta ecuación a una línea de corriente en un fluido que
cumple las hipótesis de Bernoulli, por ejemplo, en una tubería
horizontal que se angosta, tal como en la Fig.~\ref{fig1}, encontramos
que:
\begin{equation}
 P_1 + \frac{\rho v_1^2}{2} = P_2 + \frac{\rho
   v_2^2}{2},
    \label{eq2}
\end{equation}
donde $P_1$ y $v_1$ son la presión y la velocidad respectivamente en
un punto de la tubería, $P_2$ y $v_2$ las de otro punto de la tubería,
y $\rho$ la densidad del fluido. Al conjugar la ecuación
Ec.~\eqref{eq2} con la ecuación de continuidad, deducimos que cuanto
mayor es la velocidad del fluido, menor es su presión. Esta
consecuencia sólo es válida a lo largo de una tubería horizontal en un
fluido que verifica las hipótesis planteadas. Sin embargo, varios
libros de texto, no solo la destacan, sino que en la forma en la cual
la presentan, le dan un estatus de conclusión general o incluso
principio, que claramente no tiene
\cite{resnick2002physics,sears2013fisica,jewett2008fisica,tipler2004fisica}.
Seguidamente, utilizan dicho resultado para explicar un conjunto de
fenómenos tales como la sustentación de un avión o la comba de una
pelota que rota sobre sí misma. Estos interesantes fenómenos son
ampliamente conocidos y las explicaciones dadas por los textos y
artículos científicos
\cite{bauman1994interpretation,brusca1986buttressing}, rápidas y
elegantes, producen los resultados esperados. Esta \textit{facilidad
  de explicación} ha contribuido a que la premisa que mayor velocidad
implica menor presión se haya extendido en la literatura, excediendo
la ecuación de Bernoulli y sus consecuencias.

\begin{wrapfigure}{r}{0.25\textwidth}
    \includegraphics[width=0.3\textwidth]{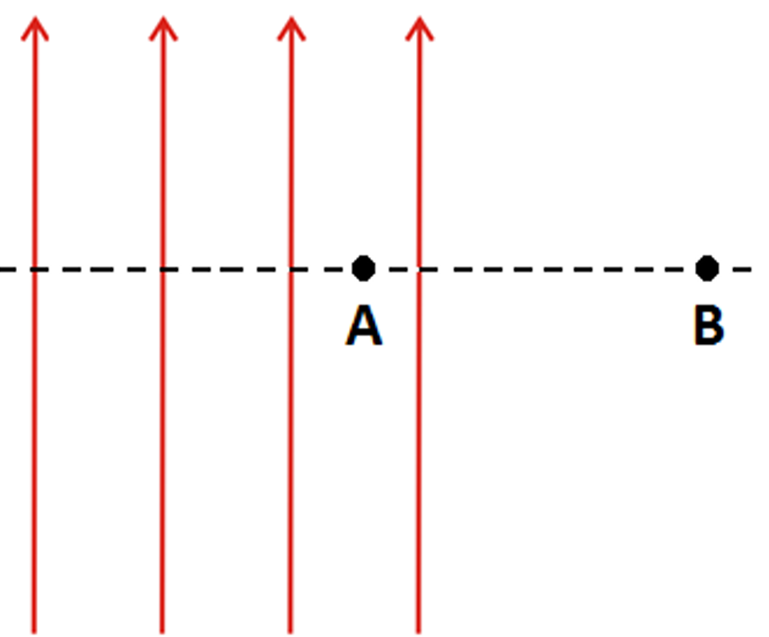}
    \caption{Chorro de fluido con velocidad uniforme. La presión en el
      punto A, interior al chorro, es la misma que en el B, exterior
      al chorro.}
    \label{fig2}
\end{wrapfigure}

Sin embargo, que la predicción de un fenómeno específico sea correcta,
no implica que su explicación también lo sea, puesto que el mismo
razonamiento aplicado en situaciones muy similares conduce a
contradicciones con los experimentos
\cite{kamela2007thinking,koumaras2018flawed}. Los fenómenos
mencionados antes como el efecto de una pelota o la sustentación de
los aviones son ejemplos de estas situaciones. Si bien podrían ser
explicados con la premisa mayor velocidad - menor presión, este mismo
razonamiento no puede explicar y conduce a contradicciones cuando
consideramos por ejemplo el efecto de la pelota en sentido contrario
al habitual conocido como efecto Magnus inverso
\cite{cross2017measurements} o el vuelo cabeza abajo de los aviones de
combate o incluso los perfiles alares simétricos en aviones
acrobáticos. En efecto, varios trabajos muestran que la sustentación
de un ala de avión, quizás el fenómeno explicado en los textos más
representativo de la ecuación de Bernoulli, no puede ser explicada a
través de una simple aplicación de dicha ecuación y sus consecuencias,
y se deben tomar en consideración otros factores
\cite{babinsky2003wings,eastwell2007bernoulli,smith1972bernoulli}. En
estos fenómenos, otros factores propios de los fluidos viscosos, como
el arrastre y el efecto Coanda, juegan un papel preponderante
\cite{eastwell2007bernoulli,smith1972bernoulli,weltner2011misinterpretations}.

Existen más situaciones en que se realiza un uso abusivo de la
ecuación de Bernoulli. Aún dejando de lado los efectos viscosos, si se
comparan las presiones en un punto dentro de un chorro de un fluido
con velocidad uniforme (punto A de la Fig.~\ref{fig2}), con la presión
de un punto exterior (punto B), dichas presiones deben ser iguales,
como muestra \citeapos{kamela2007thinking}. Este resultado, que a
priori puede resultar sorprendente, se puede entender fácilmente
aplicando las leyes de Newton. Supongamos por hipótesis que la presión
en el punto A fuera menor que en B. Si dicha consideración fuera
correcta, el gradiente de presiones provocaría una aceleración del
fluido, dejando el flujo de ser uniforme y pasando las líneas de
corriente a ser curvas. Vemos entonces, que la simple aplicación de
las leyes de Newton muestra que las presiones en A y B deben ser
iguales. La invalidez de la ecuación de Bernoulli para vincular las
presiones de los puntos A y B se debe a que por el punto B no pasa
ninguna línea de corriente, por lo tanto no es posible utilizar dicha
ecuación para comparar las presiones entre estos puntos.

Este ejemplo muestra cómo la aplicación de los principios
fundamentales de la dinámica permite comprender la cinemática del
flujo dejando además al descubierto una conclusión sacada de un mal
uso de la ecuación de Bernoulli. De igual manera, las leyes de Newton
pueden utilizarse para comprender la razón por la cual la presión es
menor en un angostamiento de una tubería. Si la velocidad del fluido
aumenta, la fuerza neta sobre un elemento de volumen ubicado en la
frontera del estrechamiento debe ser distinta de cero, por lo tanto,
la presión antes del angostamiento debe ser mayor que después.


\section{Metodología}
\subsection{Contexto de la investigación}
Con el objetivo de cuantificar el impacto que tiene el uso indebido de
la ecuación de Bernoulli respecto a las ideas previas derivadas de la
vida cotidiana y de indagar sobre la manera en que estudiantes de
cursos básicos universitarios entienden la dinámica de los fluidos,
realizamos una investigación con estudiantes de los cursos de Física
General II de las Facultades de Ciencias y de Ingeniería de la
Universidad de la República (Uruguay). Dicho curso tiene una duración
de quince semanas y una carga horaria de cinco horas semanales,
dedicando tres semanas de clase a tópicos referidos a estática y
dinámica de fluidos, siendo la bibliografía de referencia los textos
de Física General de
\citeapos{tipler2004fisica,sears2013fisica,resnick2002physics}.

El primer paso de la indagación fue la elaboración de un conjunto de
preguntas de opción múltiple. La redacción de las preguntas y sus
distractores estuvieron basados en investigaciones previas en el área,
donde asociamos las opciones incorrectas a diferentes concepciones
alternativas \cite{suarez2017students}. La primera versión considerada
adecuada fue presentada a dos especialistas que la revisaron
cuidadosamente y propusieron aportes significativos para la
mejora. Realizadas estas correcciones que permitieron validar el test
propusimos una primera instrumentación a estudiantes de la
Licenciatura en Física de la Facultad de Ciencias de la Universidad de
la República (Uruguay). Obtuvimos 27 respuestas (sobre 38 estudiantes
cursando la materia) elaboradas en forma presencial que nos
permitieron identificar algunas dificultades en la interpretación.

A partir de estas dificultades elaboramos la versión final que fue
propuesta a estudiantes de la Facultad de Ingeniería. El instrumento
consistió en 4 preguntas múltiple opción sobre diferentes
características de los fluidos ideales en contextos similares a los
problemas y ejemplos planteados en el curso y en la bibliografía de
uso corriente. Estas preguntas fueron propuestas como parte de una
prueba de carácter opcional a realizar en la plataforma web del curso
en la semana previa a la primera prueba parcial. Obtuvimos un total de
57 cuestionarios completos de cerca de 150 estudiantes matriculados en
el curso.

\subsection{Cuestionario}
El propósito de las dos primeras preguntas del test
  fue explorar la capacidad de los estudiantes para reconocer las
  interacciones de una región de fluido con su entorno. En particular,
  identificar las fuerzas ejercidas según se trate de un fluido
  confinado o libre. Las últimas dos preguntas apuntaron a la
  respuesta de los estudiantes frente a la idea de mayor
  velocidad-menor presión en dos situaciones con fluidos no
  confinados.

\subsubsection{Pregunta 1}
\begin{wrapfigure}{r}{0.275\textwidth}
    \includegraphics[width=0.15\textwidth]{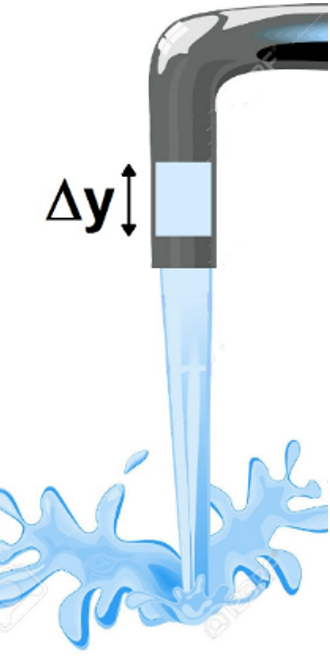}
    \caption{Cañería con elemento de fluido representado dentro de la misma.}
    \label{fig3}
\end{wrapfigure}
Considera un elemento de fluido por dentro de la cañería vertical de
altura $\Delta y$, tal como indica la Figura \ref{fig3}. De las
siguientes fuerzas en la dirección vertical:
\begin{enumerate}[1.]
    \item Una fuerza vertical hacia abajo ejercida por el líquido que
      se encuentra por encima del elemento $\Delta y$.
    \item Una fuerza vertical hacia abajo ejercida por el líquido que
      se encuentra por debajo del elemento $\Delta y$.
    \item Una fuerza vertical hacia arriba ejercida por el líquido que
      se encuentra por debajo del elemento $\Delta y$.
    \item La fuerza peso.
\end{enumerate}
¿Cuál o cuáles de dichas fuerzas actúan sobre el elemento del fluido?
\begin{enumerate}[A)]
    \item Sólo la 1.
    \item Sólo la 4.
    \item 1 y 4
    \item 1, 2 y 4.
    \item \textbf{1, 3 y 4}
\end{enumerate}
Las opciones incorrectas fueron diseñadas con los siguientes criterios:
\begin{itemize}
    \item Las opciones (a) y (d) no están asociadas a ninguna
      concepción alternativa reportada. Son distractores para reducir
      la probabilidad de acertar la respuesta azarosamente.
    \item La opción (b) refleja la idea que el fluido en la tubería
      vertical está en caída libre, es decir, que las partículas del
      fluido no interactúan entre sí.
    \item La opción (c) está asociada a la concepción alternativa que
      la presión se debe al peso de la columna por encima del elemento
      del fluido y por lo tanto la fuerza que recibe se debe al fluido
      que se encuentra encima de éste, desconociéndose que el fluido
      aguas abajo puede ejercer una fuerza.
\end{itemize}

\subsubsection{Pregunta 2}
\begin{wrapfigure}{R}{0.275\textwidth}
    \includegraphics[width=0.15\textwidth]{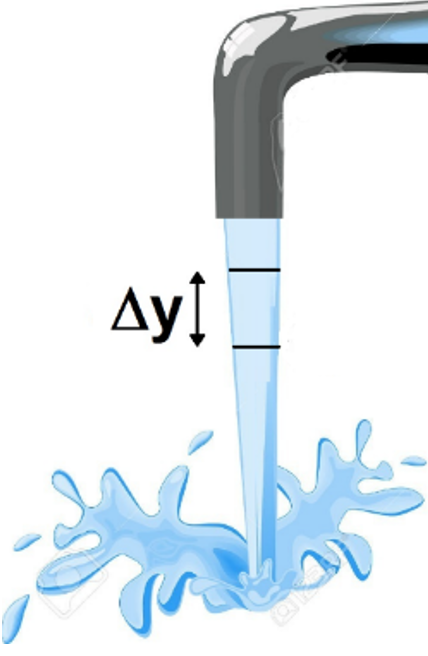}
    \caption{Cañería con elemento de fluido representado fuera de la misma.}
    \label{fig4}
\end{wrapfigure}

Considera un elemento de fluido por fuera de la cañería vertical de
altura $\Delta y$, tal como indica la figura \ref{fig4}. De las
siguientes fuerzas en la dirección vertical:
\begin{enumerate}[1.]
    \item Una fuerza vertical hacia abajo ejercida por el líquido que
      se encuentra por encima del elemento $\Delta y$.
    \item Una fuerza vertical hacia abajo ejercida por el líquido que
      se encuentra por debajo del elemento $\Delta y$.
    \item Una fuerza vertical hacia arriba ejercida por el líquido que
      se encuentra por debajo del elemento $\Delta y$.
    \item La fuerza peso.
\end{enumerate}
¿Cuál o cuáles de dichas fuerzas actúan sobre el elemento del fluido?
\begin{enumerate}[A)]
    \item Sólo la 1.
    \item \textbf{Sólo la 4.}
    \item 1 y 4
    \item 1, 2 y 4.
    \item 1, 3 y 4
\end{enumerate}

Las opciones incorrectas fueron diseñadas con los siguientes criterios:
\begin{itemize}
    \item Las opciones (a) y (d) no están asociada a ninguna
      concepción alternativa reportada. Son distractores para reducir
      la probabilidad de acertar la respuesta azarosamente.
    \item La opción (c) está asociada a la concepción alternativa que
      la presión se debe al peso de la columna por encima del elemento
      del fluido y por lo tanto la fuerza que recibe se debe al fluido
      que se encuentra encima de éste, desconociéndose que el fluido
      aguas abajo puede ejercer una fuerza.
    \item La opción (e) está asociada a la idea que las interacciones
      entre las distintas partes del fluido no dependen de si el
      fluido está o no en caída libre.
\end{itemize}

\begin{wrapfigure}{r}{0.275\textwidth}
    \includegraphics[width=0.15\textwidth]{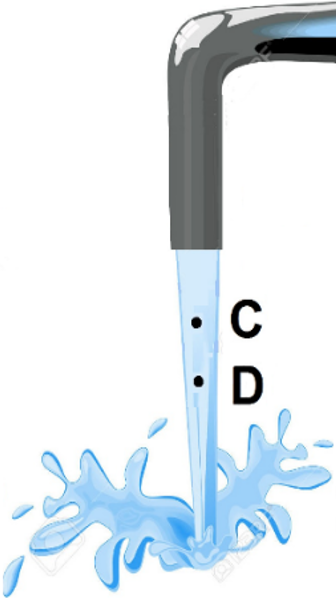}
    \caption{Chorro de agua saliendo de una cañería.}
    \label{fig5}
\end{wrapfigure}

\subsubsection{Pregunta 3}
Considera el chorro de agua después de haber salido de la cañería y
dos puntos C y D marcados en el mismo, tal como indica la figura
\ref{fig5}. Sea $P_{ATM}$ la presión atmosférica. Si comparamos las
presiones en los puntos C y D con la atmosférica, concluimos que:
\begin{enumerate}[A)]
    \item $P_C > P_D > P_{ATM}$
    \item $P_D > P_C > P_{ATM}$
    \item $\mathbf{P_c = P_D = P_{ATM}}$
    \item $P_{ATM} > P_C > P_D$
    \item $P_{ATM} > P_D > P_C$
\end{enumerate}
Las opciones incorrectas fueron diseñadas con los siguientes criterios:

\begin{itemize}
    \item La opción (a) refleja la idea que en cualquier condición la
      presión siempre disminuye al aumentar la velocidad, teniendo que
      ser la presión en chorro mayor a la atmosférica para poder
      “salir”.
    \item La opción (b) está asociada a la concepción alternativa que
      la presión se debe al peso de la columna por encima del elemento
      del fluido, así como al hecho que mayor velocidad implica mayor
      presión y que la presión en el chorro debe ser mayor que la
      atmosférica para poder “salir”.
    \item La opción (d) refleja la idea que la presión atmosférica
      “aprieta” cada vez más el chorro de agua, extrapolando además la
      idea que siempre mayor velocidad implica menor presión.
    \item Finalmente la (e) no está asociada a ninguna concepción
      alternativa previamente comunicada.
\end{itemize}

\subsubsection{Pregunta 4}
\begin{wrapfigure}{r}{0.34\textwidth}
    \includegraphics[width=0.31\textwidth]{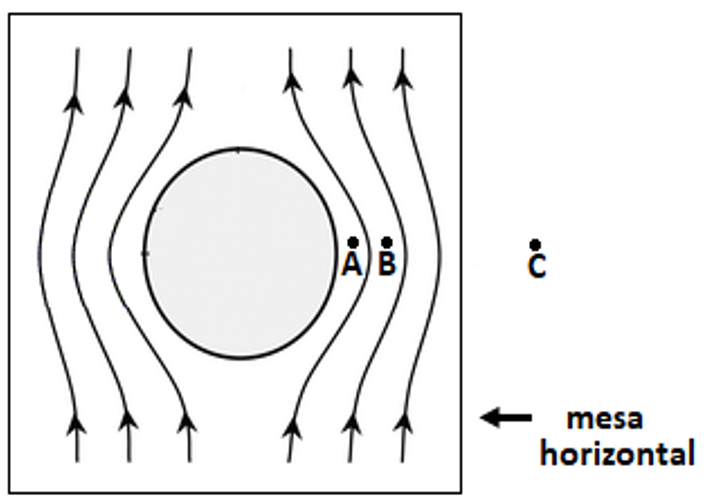}
    \caption{Líneas de corriente de un fluido que se topa con un obstáculo.}
    \label{fig6}
\end{wrapfigure}
Al pasar un fluido a gran velocidad alrededor de un cilindro apoyado
sobre una mesa horizontal, las líneas de corriente se deforman,
adquiriendo la forma indicada en la figura \ref{fig6}.  Los puntos A,
B y C se ubican a igual altura. El punto C está alejado de las líneas
de corriente, encontrándose a la presión atmosférica $P_{ATM}$. En los
puntos A y B las líneas de corriente realizan arcos de circunferencia
con aproximadamente la misma velocidad.  Si comparamos las presiones P
en los puntos A y B con la presión en C ($P_C=P_{ATM}$), concluimos
que:

\begin{enumerate} [A)]
    \item $P_A = P_B = P_C$
    \item $P_A > P_B > P_C$
    \item $P_A = P_B > P_C$
    \item $P_C > P_B = P_A$
    \item $\mathbf{P_C > P_B > P_A}$
\end{enumerate}

Las opciones incorrectas fueron diseñadas con los siguientes criterios:
\begin{itemize}
    \item La opción (a) refleja la concepción alternativa que a igual
      altura igual presión.
    \item La opción (b) no está asociada a ninguna concepción
      alternativa conocida.
    \item La opción (c) refleja la idea previa que mayor velocidad
      implica mayor presión.
    \item Finalmente la (d) refleja la concepción alternativa que
      siempre mayor velocidad implica menor presión.
\end{itemize}


\section{Resultados}
\begin{wrapfigure}{r}{0.3\textwidth}
    \includegraphics[width=0.35\textwidth]{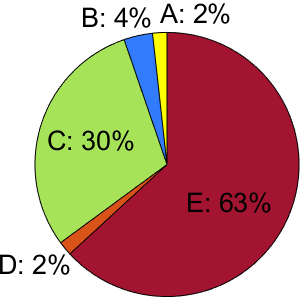}
    \caption{Respuestas a la pregunta 1.}
    \label{fig7}
\end{wrapfigure}

\begin{wrapfigure}{r}{0.3\textwidth}
    \includegraphics[width=0.3\textwidth]{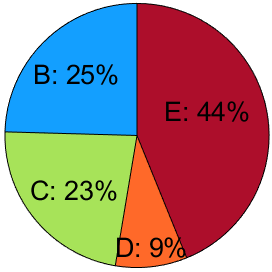}
    \caption{Respuestas a la pregunta 2.}
    \label{fig8}
\end{wrapfigure}

En la primera pregunta, aproximadamente uno de cada tres estudiantes
se inclinó por el distractor más fuerte (ver Fig.~\ref{fig7}),
considerar que las únicas fuerzas que actuaban sobre el elemento del
fluido eran el peso y la debida al líquido que se encontraba por
encima del fluido, e ignorar que el fluido aguas abajo ejercía una
fuerza. Mientras que, en la segunda, aproximadamente dos de cada tres
estudiantes no fueron capaces de reconocer que el elemento del fluido
se encontraba en caída libre (ver Fig.~\ref{fig8}), considerando que
actuaban además del peso, una fuerza ejercida hacia abajo debida al
fluido que se encontraba por encima del elemento y en algunos casos
también una ejercida por el fluido aguas abajo, vertical hacia arriba.

Las respuestas a estas preguntas permitieron verificar una de las
dificultades conceptuales más importantes relevadas en la
literatura. El hecho que los estudiantes presentan grandes
dificultades para reconocer la manera en que interactúan las distintas
partes del fluido, no siendo capaces de conectar la cinemática con la
dinámica. Esto surge de ambas preguntas. En la primera, donde no
reconocían que el fluido debía moverse con velocidad constante dentro
de la cañería, o en caso de hacerlo, no eran capaces de
compatibilizarlo con el principio fundamental de la dinámica y
principalmente en la segunda, donde el fluido estaba en caída libre y
sin embargo la mayoría de los estudiantes consideraban que actuaban
otras fuerzas sobre el elemento además del peso.

Al analizar en forma global las respuestas a ambas preguntas, vemos
que los estudiantes tuvieron grandes dificultades para reconocer la
diferencia entre ambas situaciones (el elemento del fluido confinado y
no confinado). Por lo tanto, no han elaborado un modelo adecuado sobre
como interactúa un elemento de un fluido con su entorno.

\begin{figure}[h!]
    \centering
    \begin{subfigure} [h]{0.45\textwidth} 
        \includegraphics[scale=0.7]{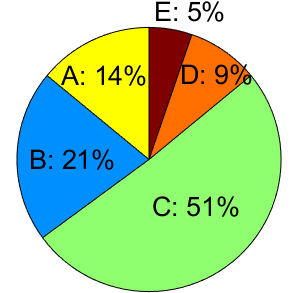}
        \caption{Respuestas a la pregunta 3.}
        \label{fig9}
    \end{subfigure}
    \begin{subfigure} [h]{0.45\textwidth} 
        \includegraphics[scale=0.7]{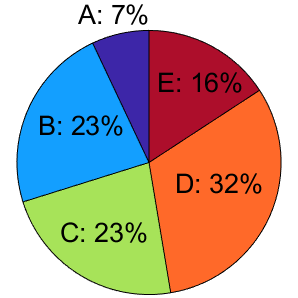}
        \caption{Respuestas a la pregunta 4.}
        \label{fig10}
    \end{subfigure}
    \caption{Respuestas a las preguntas 3 y 4}
\end{figure}

En la tercera pregunta, los estudiantes tenían que comparar las
presiones en diferentes puntos del fluido no confinado que se
encontraba en caída libre a presión atmosférica. Del análisis de los
resultados se desprende que la mitad de los estudiantes no fueron
capaces de reconocer que la presión de todos los puntos dentro del
fluido debía ser la atmosférica (ver Fig.~\ref{fig9}). A su vez, de
las respuestas incorrectas, la mitad de ellas estaba asociada a la
extrapolación de la idea que mayor velocidad implica una menor
presión, mientras que la otra mitad a la idea contraria. La cuarta
pregunta presentaba nuevamente una situación donde se encontraba un
fluido no confinado, pero en este caso, correspondía a un fluido que
se movía con una velocidad de módulo aproximadamente constante, pero
sus líneas de corriente se deformaban por la presencia de un
obstáculo. En esta pregunta, los estudiantes debían comparar las
presiones entre puntos del fluido cada vez más alejados del
obstáculo. Para esta cuestión 4 de cada 5 estudiantes no fueron
capaces de reconocer que la presión debía aumentar a medida que uno se
alejaba del obstáculo, para proporcionar la fuerza neta centrípeta
necesaria para que un elemento del fluido realizara un movimiento
curvilíneo (ver Fig.~\ref{fig10}). En esta pregunta, los distractores
estaban asociados a las mismas concepciones alternativas que en la
cuestión anterior, destacándose que de las respuestas incorrectas 2 de
cada 5 estaban asociadas nuevamente a la idea que mayor presión
implicaba menor velocidad.

Del análisis de los resultados de las últimas dos preguntas, queda
claro que muchos estudiantes después de haber visto en el curso de
Física los conceptos básicos de la dinámica de fluidos ideales,
extrapolan la idea que mayor velocidad implica una menor presión en
contextos donde no es válida. Es interesante notar que estas preguntas
se pueden contestar correctamente aplicando el principio fundamental
de la dinámica a un elemento del fluido. Esto deja al descubierto la
idea ya sugerida en la literatura
\cite{suarez2017students,schafle2019students} que los estudiantes
tienen grandes dificultades para comprender la manera que interactúan
las distintas partes de un fluido.

\section{Comentarios finales}
La investigación realizada muestra que aún después de haber transitado
por cursos de Física General un conjunto importante de estudiantes
mantiene la idea previa, originada de la experiencia cotidiana, que
una mayor velocidad en un fluido implica una mayor
presión. Similarmente, un conjunto semejante evidenció la concepción
alternativa que mayor presión siempre implica una menor velocidad
independientemente del contexto. De esta manera, parece claro el
impacto negativo que tiene en las ideas de los estudiantes los usos
abusivos frecuentes en muchos libros de texto, así como material de
divulgación, de la ecuación de Bernoulli. En este sentido, la
concepción mencionada es un producto del propio sistema educativo.

En paralelo y consonancia con las últimas investigaciones en el área,
se ha verificado que muchos estudiantes aún no han elaborado un modelo
adecuado de la interacción de un elemento de un fluido con su
entorno. Este hecho es fundamental, ya que las cuatro preguntas
planteadas, pueden contestarse correctamente aplicando el principio
fundamental de la dinámica sin tener un conocimiento de la ecuación de
Bernoulli. En ese sentido, creemos firmemente que si en los libros de
texto y en los cursos se hiciera un análisis dinámico de las fuerzas
que actúan sobre un elemento del fluido, se podría lograr que los
estudiantes elaboren modelos más sofisticados de la mecánica de los
fluidos. Este cambio de enfoque permitiría a los estudiantes tener una
imagen correcta de diversos fenómenos (como por qué vuelan los aviones
o el efecto en una pelota de fútbol) sin necesidad de hacer
aproximaciones erróneas ni involucrar conceptos de una complejidad
superior a la que ya se tiene en los cursos básicos de dinámica de
fluidos.

En la ensen\~anza de la ciencias se acepta
  generalmente que adem\'as de detectar las concepciones alternativas
  deber\'\i amos ser capaces de responder de d\'onde proceden y sobre
  qu\'e factores debemos incidir para favorecer un cambio conceptual
  \cite{benarroch2001interpretacion}. Se ha propuesto en la literatura
  que en ciertas ocasiones las concepciones alternativas son
  originadas por errores u omisiones en los libros de texto, causadas
  muchas veces por el hecho que no se realiza un esfuerzo por
  evitarlas o se lo hace en forma deficiente, o simplemente porque los
  propios libros tienen errores conceptuales graves
  \cite{carrascosa2005problema}. En este sentido, determinar si las
  concepciones alternativas aqu\'{\i} discutidas se originan en ideas
  previas provenientes de la experiencia cotidiana, anteriores al
  tránsito por el sistema educativo, o por el contrario provienen del
  propio sistema, es un problema que permanece abierto para futuras
  investigaciones.

Destacamos que resulta conveniente presentar a los estudiantes
actividades experimentales desafiantes y que pongan en juego sus
concepciones alternativas. En la literatura se han presentado diversos
experimentos que muestran las contradicciones de extrapolar la idea de
mayor presión-menor velocidad a contextos donde no es válida. Creemos
que la actividad experimental sugerida por
\citeapos{dutra2020quarter}, adaptación del experimento de
\citeapos{ehrlich1990turning} donde se sopla una moneda hasta
levantarse, pero con la moneda en una ranura ligeramente más grande
que ella de manera que quede al ras de la superficie, al ser muy
sencillo y de bajo costo, puede replicarse fácilmente en cualquier
aula, utilizándose en una modalidad del tipo POE
\cite{tenreiro2006diseno}. Este tipo de experiencias se podrían
utilizar en museos de ciencias, donde es necesario incorporar
materiales didácticos como los desarrollados por
\citeapos{guisasola2005diseno} para mejorar el aprendizaje de los
estudiantes en una visita. Finalmente, creemos que es importante
indagar en otras poblaciones y en diferentes niveles educativos el
peso de las concepciones alternativas en fluidos surgidas del propio
sistema educativo, así como sobre las ideas de los docentes vinculadas
a las mismas.

\nocite{*}

\bibliographystyle{apacite}
\bibliography{mybiblio}

\end{document}

%% file: msel.tex
\usepackage{theorem}
\theoremstyle{plain}{

}

\theoremstyle{plain}{

}
\usepackage{fancyhdr}
\renewcommand{\sectionmark}[1]{\markright{{\sf   {\sl Volume $\bullet$}, $\bullet$. }}{}}
\usepackage[Lenny]{fncychapmsel}
\usepackage{amssymb,amsmath,amsfonts}
\usepackage{enumerate}
\usepackage[greek,spanish,english]{babel}
\usepackage[mathscr]{eucal}
\usepackage{color}
\usepackage{graphicx}
\usepackage{booktabs}
\usepackage{minitoc}
\def\abst#1{
\begin{center}
\begin{minipage}{17cm} \it
\hrule
\begin{center}
$\vspace{0.2cm}$\\
  {\bf Abstract}\\
\end{center}
\begin{center}
\begin{minipage}{16cm}
  #1
\end{minipage}
\end{center}
\hrule

\end{minipage}
\end{center}}
\definecolor{Azulon}{rgb}{0.65,0.75,1}
\newcommand\logoMSEL{
\begin{minipage}{4cm}
\resizebox{4cm}{!}{\textcolor{blue}{ \fontsize{14}{16}\usefont{OT1}{pag}{m}{n}\selectfont \bf \color{Azulon}{MSEL}}}
\resizebox{4cm}{2mm}{\sc \hspace{2.5cm} \rotatebox{90}{\hspace{-1.8cm} Modelling} in Science Education and Learning}
\end{minipage}}

\newenvironment{keyword}{{\sf Keywords: }}%
{\endtrivlist\medskip}

\newenvironment{clave}{{\sf Palabras clave: }}%
{\endtrivlist\medskip}

{\endtrivlist\medskip}

{\endtrivlist\medskip}

\usepackage{hyperref}
\hypersetup{pdfpagemode=UseOutlines,colorlinks=true,citecolor=blue}
\usepackage{apacite}

\makeatletter
   \def\cleardoublepage{\clearpage\if@twoside \ifodd\c@page\else
  \vspace*{\fill}
     \thispagestyle{empty}
    \newpage
    \if@twocolumn\hbox{}\newpage\fi\fi\fi}
\makeatother

\def\vol{$\bullet$}
\def\anyo{$\bullet$}